# Dynamic Program Slices Change How Developers Diagnose Gradual Run-Time Type Errors

Felipe Bañados Schwerter[a], Ronald Garcia[b], Reid Holmes[b], and Karim Ali[c]

a University of Alberta, Canada

b The University of British Columbia, Canada

c NYU Abu Dhabi, UAE

**Abstract** A gradual type system allows developers to declare certain types to be enforced by the compiler (i.e., statically typed), while leaving other types to be enforced via runtime checks (i.e., dynamically typed). When runtime checks fail, debugging gradually typed programs becomes cumbersome, because these failures may arise far from the original point where an inconsistent type assumption is made. To ease this burden on developers, some gradually typed languages produce a blame report for a given type inconsistency. However, these reports are sometimes misleading, because they might point to program points that do not need to be changed to stop the error.

To overcome the limitations of blame reports, we propose using dynamic program slicing as an alternative approach to help programmers debug run-time type errors. We describe a proof-of-concept for TypeSlicer, a tool that would present dynamic program slices to developers when a runtime check fails. We performed a Wizard-of-Oz user study to investigate how developers respond to dynamic program slices through a set of simulated interactions with TypeScript programs. This formative study shows that developers can understand and apply dynamic slice information to provide change recommendations when debugging runtime type errors.

## 1 Introduction

When software developers choose a programming language for their project, they are also choosing to follow its type system. Trade-offs in the latter choice impact the software development process. If developers choose a statically typed language (e.g., Rust), their deployed code is guaranteed to be free of certain categories of errors (e.g., broken type interfaces, memory safety), depending on the provable soundness properties of the specific type system.

However, these guarantees come at a cost for developers, who must commit to type interfaces early in the development process and appease the type checker before they can run their programs. This constraint hinders prototyping, when developers want to (but cannot) run incomplete pieces of software. Alternatively, if developers choose a dynamically typed language (e.g., JavaScript), prototyping is easier, but the final product has fewer guarantees about potential run-time errors. Gradual typing [31, 35] is an intermediate approach to types, easing software evolution by allowing developers to choose where and when to commit to static type checking in their project, at their own discretion and pace. With gradual typing, programmers can stay in the same programming language throughout the lifetime of a project and migrate from the prototyping benefits of dynamically typed languages towards the guarantees of statically typed ones. Gradual typing has inspired the development of several languages such as Typed Racket [25, 36], TypeScript [23], Hack [12], and the Mypy [33] checker for Python.

Programmers in gradually typed languages expect that the type declarations they provide are respected when programs run [37], and expect that the language runtime will produce errors when they are not. In the presence of function types (e.g., to assign types to callback parameters), required checks cannot be immediately resolved and must be wrapped around function values to check each argument when the function is called and the corresponding result after a function call returns. In some cases, these extra checks may impose significant runtime overhead when executing programs [32], and complicate the generation of debugging error messages when a check fails [30]. Because of the performance overhead of these run-time checks, the designers of TypeScript chose to skip them altogether. This choice leads to unexpected program behaviour (type unsoundness), triggering errors that may point toward parts of a program irrelevant to the failure at hand.

Unlike TypeScript, Typed Racket [36] is guaranteed to introduce sufficient run-time checks, and includes a run-time error reporting system based on the semantics of *blame* [10, 11, 15, 38]. Blame errors point towards the place in a program declaring a type assumption that the program later breaks. However, blame reports also assume that all type annotations are correct, which biases error messages [8]. For some kinds of bugs, blame reports are misleading. Therefore, researchers have proposed strategies that developers may follow to use blame reports to triangulate the causes of a run-time error [20, 21]. However, those strategies do not always work. Thus state-of-the-art error reports for run-time check failures in gradually typed languages do not always help developers in finding the causes of the failures that the runtime detects.

As a complementary approach, we propose applying dynamic program slicing [19, 24, 28, 39] to gradually typed languages. Program slicing distinguishes parts of a program that are provably irrelevant to the computation of a specific result. In the case of a gradual run-time error, the failure is the result, and we can use backward program slicing to dynamically distinguish irrelevant type constraints in the original program and avoid unnecessary distractions during debugging. To the best of our knowledge, we are the first to propose using slicing systems as a technique diagnosing gradual type errors. In this paper, we investigate the hypothesis that this approach will help developers debug and understand programs better.

The design of a semantics for slicing systems for gradually typed languages is an interesting formal problem [1], involving mapping backwards through type-directed compilation from the instrumented intermediate runtime language so that error reports can be faithfully presented against source programs. However, any formal justification must also consider the empirical dimensions of such a system. To that end, this paper does not focus on the formal aspects of designing a slicing system, but rather on empirical questions like “Can developers use a program slicing system to identify causes of run-time type failures?”, “Will they use it?”, “Does such a system help developers?”, or “Is the information in the slice consistent with developer expectations?”. To collect insights related to these questions, we have developed a prototype interface for *TypeSlicer*, a dynamic program slicer for gradually typed programming languages. We performed a user study to collect qualitative feedback from real-world developers on the use of TypeSlicer as a debugging aid in the presence of run-time type errors. To avoid the considerable engineering effort required to build a slicing system for a real-world programming language, we follow the Wizard-of-Oz [18] methodology. This type of experiment is intrinsically limited, because it does not allow participants to freely alter the programs that they debug or try their own programs. Nevertheless, the methodology allows us to evaluate program slicing in the context of a general-purpose language with a large participant pool of real-world programmers, before incurring the cost of developing a general-purpose real-world slicing system.

Our user study with 30 participants shows that 30 % of participants relied solely on the available program slice to understand the bug and provide a change recommendation. For many tasks, participants went back and forth from the slice to the complete program, while only 11 % turned off the slice quickly and likely disregarded the slice information. Overall, participants provided positive qualitative feedback and on average gave 75/100 points to the TypeSlicer interface on the standardized System Usability Scale [5]. The artifact associated with this paper is available at [3].

**Program Slicing at a High Level.** A program slice is a selection of portions of a program. Portions which are included in the slice may be relevant to compute the program result, whereas all portions missing from the slice must be irrelevant. For the purposes of this paper, results of interest are runtime errors that we intend to debug. Originally [39], program slices were just sets of line numbers calculated as a static analysis, but the approach has evolved to encompass arbitrary portions of programs and dynamic analysis [19, 24, 28], allowing slices to be more precise and identify wider parts of a program as irrelevant, and allowing the analysis to act over arbitrary portions of both the final result and of intermediate program states.

To obtain program slices from an execution trace, one needs some specific slicing algorithm. We work with an extension of Galois slicing [28], an approach that requires a forward slicing (or trace replay) and a backward slicing (or trace rewinding) algorithm. Intuitively, when parts of a program are sliced out, the recorded trace can either reproduce the output or run out of information to recreate the failure. Forward slicing provides a trace-based semantics for program slices, where the code marked as irrelevant in a slice is propagated through the trace to obtain slices of later program states, under the intuition that any changes to the redacted parts of the code would not make the errors go away (though they may introduce new errors that preempt the current ones).

For example, Program 5 in Listing 3 runs to an error: inside the `process` function, `y` is expected to be an array of numbers, but in this case the function was called with an array of strings. Since there is no type annotation on the function parameter, the gradual type system cannot see the inconsistency statically, it is only detected at runtime. The slicing system aims to produce the smallest slice that yet still reproduces the failure via trace replay, with the intent of reducing the attention burden of a debugging developer. For example, changing the code *after* the call to process, where the failure is revealed, would still reproduce the failure. Therefore any slice marking that code as irrelevant should still reproduce the failure. In our example program, given that the problem happens at the type level, we don't need to keep the value assigned to `x`, but only the places where the type inconsistencies arise: the assignment of the parameter as an array of numbers, the declaration of `x` as an array of strings and of `y` as having an unknown type, and the function call.

## 2 The TypeSlicer Graphical User Interface

TypeSlicer is a fantasy name for a program slicing tool for TypeScript. In our study, we present participants with an on-line programming interface extended with mock-up slices for a predetermined set of example programs. We chose TypeScript for its popularity as a "gradual" language, in spite of its safety limitations. The optional type annotations in TypeScript do not introduce any extra type checks when the program runs, although safe versions of Typescript have been proposed [27]. The abundance of TypeScript programmers considerably widens the candidate population for our study. We describe TypeSlicer to participants as a type-safe runtime semantics that addresses the safety limitations of TypeScript and detects runtime type inconsistencies. To debug these inconsistencies, TypeSlicer provides program slices: parts of the program that are guaranteed to not be involved in the error are blurred out.

Figure 1 shows the online programming interface that we present to our study participants. When a program runs into a type inconsistency, TypeSlicer automatically blurs out the portions of the program that a slice guarantees to not be involved in the failure. We prototyped TypeSlicer using the same editor [22] that also powers Visual Studio Code. This editor includes TypeScript syntax highlighting and integrates

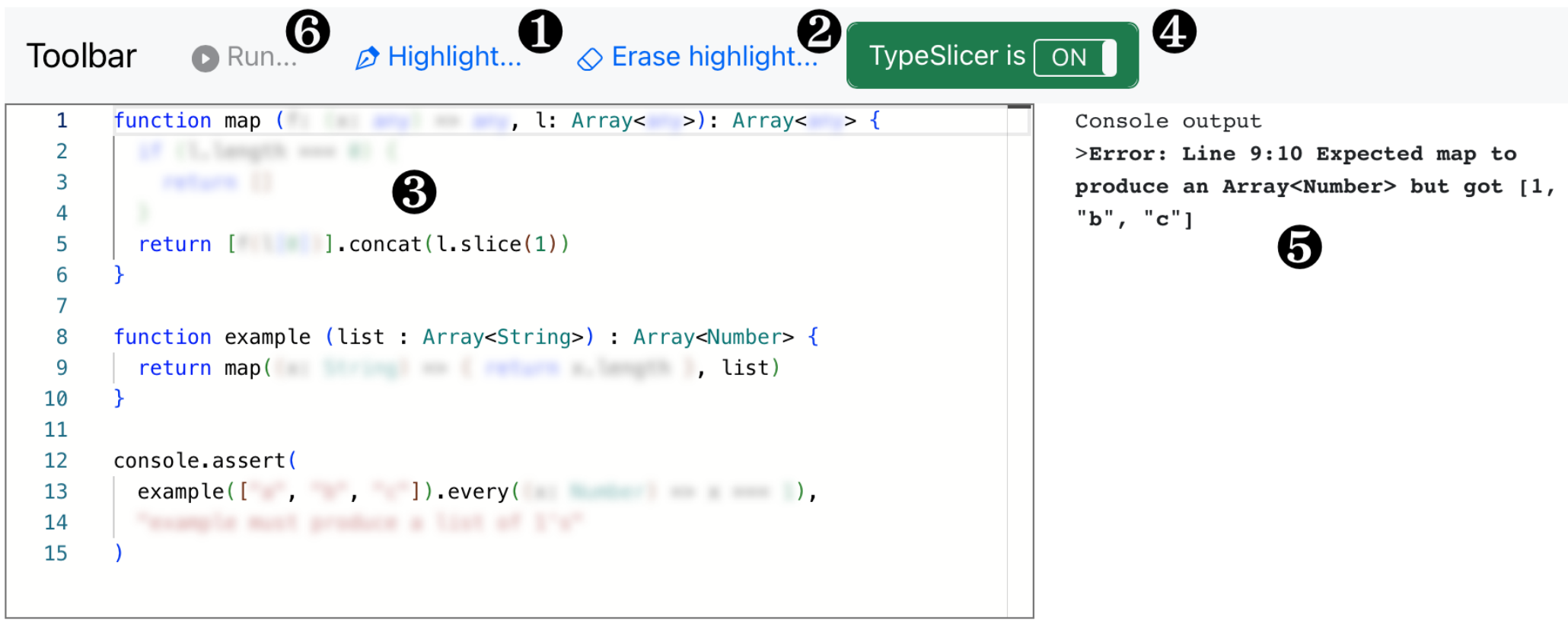

■ **Figure 1** The graphical user interface for TypeSlicer.

a TypeScript compiler, providing access to type checking, compilation, and type information from the web browser. We extended the base text editor to include:

- The option for participants to highlight multiple parts of the program with the button in ❶. These highlights can also be removed with the button in ❷.
- A presentation layer for program slices on top of the code. Whatever is marked in the slice as irrelevant appears in the interface as blurred text, as seen in ❸. When the slicing tool is available, a toggle button seen in ❹ allows participants to hide the slice and see the full program text.
- A console output viewer, seen in ❺, presents some text error message when participants click on the Run button seen in ❻. Once this button is clicked, it deactivates itself for purposes of the study.

We chose to blur the code marked as irrelevant in the slice to provide a strict distinction between moments when participants look at the slice information and when they may look at the program context. Toggling the tool on or off acts as a proxy for switching attention between the slice and the rest of the program.

Due to the limitations of the Wizard-of-Oz approach, participants were not allowed to change the code of the program. This restriction also ensured that participant highlights remained comparable. Participants also retained access to the feedback tools from the underlying editor. When participants hover over an identifier, they can see the type that TypeScript assigns to it. When they click on an identifier, the editor highlights all other references to the name in the rest of the program.

The program slice and the console output constitute the Wizard-of-Oz portion of the experiment and are provided by the experimenter, not directly computed by the tool deployed in the experiment. We computed the slices presented in the experiment in a simpler language calculus and then manually translated them to equivalent TypeScript programs. Although not the focus of this paper, we have developed a formal framework for the development of slicing systems for gradual languages that we used to generate the slices in the study [1]. To generate the slices, we developed equivalent programs in $\text{GTFL}_{\lesssim}$ [2], a gradual language with record subtyping designed using Abstract

Gradual Typing (AGT) [14]. We follow the semantics of $GTFL_{\lesssim}$ to compute program slices which we then manually translate back into TypeScript syntax.

## 3 Research Questions

The goal of our user study is to investigate whether programmers can use program slices to debug runtime type errors. This question may only be answered indirectly, because programmers are not directly exposed to a formally defined program slice. Instead, the program slice is interpreted and presented through the TypeSlicer user interface. If participants react positively to TypeSlicer, they are also reacting positively to the slices being presented. But if participants were to react negatively to TypeSlicer, we cannot know if they are reacting negatively to slices in general, as they might still react positively to a different (unexplored) presentation of program slices.

Throughout our study, we focus on three related research questions: RQ1 *Will developers use our slicing interface design to identify potential causes of runtime type failures?* This question aims at empirically supporting the hypothesis that developers will try to use slices if available for gradual programs. We explore the use of slices as debugging aids, so developers will try to identify causes of failures and propose fixes or changes to the code. RQ1 is related to two other secondary questions: RQ2 *Do developers find slices useful?*, and RQ3 *Do developers rely on slice information to fix program errors?* Both RQ2 and RQ3 hint at whether developers would trust and use slices for debugging. If developers perceive slices as useful and rely on their information, they are more likely to use a full-fledged implementation of TypeSlicer in the future.

## 4 User Study Design

We performed an online formative randomized user study, using Prolific [26] as a recruiting platform. This online platform acts as an intermediary for research studies, managing payments and providing limited ways to screen for participants.

First, participants answered a pre-screening questionnaire. Those successful in the pre-screening were then invited to the main study.

### 4.1 Pre-screening Questionnaire

To be included in our study, a participant needed to have worked in the software industry and have familiarity with TypeScript. The platform could screen only for the first criterion, so we designed a 3 question pre-screening questionnaire to check for TypeScript familiarity. The questionnaire takes no longer than 10 minutes to finish, but we did not enforce this time limit. The questionnaire first asks candidates to list programming languages in which they have written at least 1000 LOC (Q1). If they are familiar with TypeScript, then we provide 4 similar programs: One in Python, one in C#, one in TypeScript and one in Swift, and ask to identify all valid TypeScript

**Listing 1** Tutorial program, highlighting the lack of type safety in TypeScript. The program type checks but runs to a run-time type error.

```
var person: {id:string} = { id: "hello" }
var base_id: number = 1
function add_id (x: {}) : { id? : number } {
  let ans : { id? : number } = x
  ans.id = base_id++
  return ans
}

var z = add_id(person)

console.log(person.id.toUpperCase())
```

programs (Q2). If they successfully identify the TypeScript program alone, we then provide them with a second TypeScript program and a list of statements about its type constraints —for example, "person has type string", "y has type (number | string)[]", etc.— and we ask them to select all the type constraint statements that correctly apply to the program (Q3).

Since the questionnaire does not provide any tool support, which is usually available to TypeScript developers, several participants did not completely detect all type constraints or were confused by subtleties in the semantics of TypeScript, as we discuss in Section 6.

### 4.2 Main User Study

The main study consists of three parts: a tutorial, a set of 6 debugging tasks, and an exit questionnaire.

#### 4.2.1 Tutorial

To ensure that participants met the technical requirements for the main study, we asked them to first complete a brief, fully guided tutorial task (10 minutes). The tutorial introduced TypeSlicer and guided participants through highlighting tasks, similar to the ones that they later faced in the study. It also served as an opportunity to remind participants of the lack of type safety in TypeScript, by introducing them to a program that has type inconsistencies that the compiler does not detect (Listing 1), leading to program failure that a safe type system would detect and report statically. We presented TypeSlicer to participants as providing two main features:

- tracking type information to check for inconsistencies throughout program evaluation, providing type safety using sound gradual typing instead of optional typing.
- collecting and analyzing an evaluation trace to identify program slices that are not involved in a detected type contradiction, and hiding them in the user interface.

We frame the blurring in slices as "hiding parts guaranteed to not be involved" because that is the property our slicing framework can guarantee. Programmers would likely prefer a system "keeping parts involved in the failure" only, a related property, but the precision constraints on our analysis cannot guarantee that the slices preserve the latter, only the former.

#### 4.2.2 Debugging Tasks

We presented each participant with 6 debugging tasks. For each task, we presented each participant with a text editor that contains a program to debug in the TypeSlicer interface. The program compiles successfully, but runs to an error. A console window to the right of the interface (see ❺ in Figure 1) presents the error output when the user runs the underlying program by clicking the "Run" button ❻. Clicking the button only simulates the run, providing instead a stored error message and a stored program slice, acting as the "Wizard-of-Oz" portion of our experiment. In the toolbar, participants can also access a green toggle ❹ for TypeSlicer when available, and highlighting facilities after the program is run (❶ and ❷). The task description lists 3 actions to perform:

- Run the code, think of candidate fixes for this error, and highlight the parts of the program that need to be changed to fix the bug. All highlights were recorded locally in the participant's browser and submitted to our research server after the experiment was finished.
- To clarify why participants highlighted specific program fragments, the task description asks participants to report their understanding of the program's issues by writing a change recommendation that would be attached to a bug report of the failure.
- Write any comments about the program behaviour and whether anything was surprising or unexpected.

Participants were told to imagine themselves as recently joining a team developing a TypeScript application and being assigned to debug some errors. We mentioned that the owner of the bug reports would welcome any suggestions on how to fix the bugs presented. Besides these instructions, participants were free to select any parts of the program they chose to mark as requiring a change. Participants did not need to provide new code, but we asked them to describe a possible solution to the presented programming bugs, reflecting their understanding of the program.

We expected each debugging task to take approximately 5–10 minutes. We required participants to move on to the next task at the 15 minute mark. All participants were shown the same set of 6 programs in a randomized order to mitigate the impacts of participant fatigue throughout the experiment. Participants had access to TypeSlicer in 3 programs. The other 3 programs act as our control setting, where participants did not have access to TypeSlicer. We also separately randomized the programs for which we made TypeSlicer available for each participant.

#### 4.2.3 Exit Questionnaire

Participants were given the opportunity to provide comments about TypeSlicer, the programs, or the errors that they faced throughout the debugging tasks. At the end of the experiment, we asked participants to answer a standard usability survey (the System Usability Scale [5]).

The survey measured whether usability issues with TypeSlicer itself hampered participants, as opposed to the information available from the program slices. This usability scale acts as a baseline against which to measure bad usability of an interface [6], and its scores provide additional feedback about the TypeSlicer interface that we have designed.

## 5 Selection of Subject Programs

In picking and designing the programs for this user study, we could have focused on very simple programs with a single issue and a single solution. By doing so, we could aim to evaluate directly whether slicing helps developers find "the" problem in their code. But if the bug is simple enough, it is likely that (good) programmers would not need any tool to help them fix the problem.

The standard approach to debug sound gradual programs is the use of blame reports generated by a blame semantics. Researchers have shown that these blame reports do not always work, with some example erroneous programs that are complex enough to surface issues with these blame reports: for some programs, blame error messages point to misleading places that do not need to be changed to address the program failure [21]. We want to include in our program selection some of these errors from the literature, where alternative approaches are likely to provide either confusing or misleading information. Can slicing provide an extra debugging tool for those situations where other approaches do not work?

We adapted some example programs from related literature on blame reports [21] and gradual typing evaluation [37], aiming to keep their interesting parts while still allowing for participants to finish each task in the allotted time. We also added some new small programming problems. We aimed for each subject program to be self contained and possible to debug in a 5–10 minute debugging session. We validated the length of these debugging sessions while prototyping the user study.

This process resulted in selecting 6 programs for our study, representing runtime type errors with different levels of complexity. We discuss only four of them in the main paper. Program 2 and Program 4 are included in the appendix for space reasons and since the insights from them also arise on the other programs.

### 5.1 Program 1: Wrong-Map

Listing 2 presents Program 1, where `example()` relies on an ill-implemented `map()` function (Line 9) that applies its functional argument `f` only to the first element of the input array (Line 5). While the type signature of `map()` is never broken, calling the function assumes that it completely transforms its input into an array of numbers, the type of the range of `f`. However, the returned array is not of just numbers but instead of numbers or strings.

We selected Program 1 because the type inconsistency arises in the body of `example()`, although the implementation issue is in `map()`. This program is a simplification of an

**Listing 2** The TypeScript source code for Program 1.

```
function map (f: (x: any) => any, l: Array<any>): Array<any> {
 if (l.length === 0) {
  return []
 }
  return [f(l[0])].concat(l.slice(1))
}

function example (list : Array<String>) : Array<Number> {
  return map((x: String) => { return x.length }, list)
}

console.assert(
  example(["a", "b", "c"]).every((x: Number) => x === 1),
  "example must produce a list of 1's"
  )
```

**Listing 3** The TypeScript source code for Program 5.

```
1 function process(y): void {
2   let z: Array<number> = y
3  z.unshift(y.length)
4 }
5
6 let x: Array<string> = ["hello", "bye"]
7 process(x)
8 console.assert(
9  x[0] === "2",
10  "List includes its length at the head"
11 )
```

example described in the analysis of blame error reports for a mutation of a program that implements an infinite stream of prime numbers [21].

The highlights in gray in Listing 2 represent the portions of the program marked as irrelevant by our slicing process and which would be blurred out by the TypeSlicer interface. We present similar highlights for every program listing in this section.

### 5.2 Program 5: Push Element

Listing 3 presents Program 5, where a single type inconsistency crosses through an `any` type annotation. After the call to `process()` (Line 7), whose argument `y` is of type `any`, the contents in the list `x` are not of a uniform type.

Program 5 is a variant of a simple program in the literature [37], with a slightly more complex control flow. We selected this program as a simple example where the type inconsistency is not immediately apparent when the programmer reads the code top to bottom. The `any` crossing happens at the parameter of `process()`. The parameter

**Listing 4** The TypeScript source code for Program 3.

```
1 let hide = function(x) { return x }
2
3 let strings = hide(["-1", "0", "1", 2, "3", "4", "5"])
4
5 function next_string() {
6   return strings.pop()
7 }
8
9 function strings_only(x: string): void {
10  console.log( x.toUpperCase() )
11 }
12
13 function asks_for_2_strings(){
14   for (var i = 0; i <= 2; ++i) {
15     strings_only( next_string() )
16   }
17 }
18
19 function asks_for_1_string(){
20   strings_only( next_string() )
21 }
22
23 asks_for_2_strings()
24 asks_for_1_string()
```

is an array of strings, but the array is then used in the function at an inconsistent type by pushing into it a new value, the number y.length (Line 3).

### 5.3 Program 3: 3-Strings

Listing 4 presents Program 3, where an incorrect bound in the for loop in asks_for_2_strings() (Line 14) extracts 3 strings instead of 2. This issue is not immediate though, because the list actually has 3 strings at the beginning, but forces the next call to asks_for_1_string (Line 24) to fail even though the method itself has no errors. The type system cannot see the error immediately since the hide function masks the type of strings to be any (Line 3).

Program 3 reflects a failure caused by an error that is not in the immediate evaluation stack. This is a simplification of a program used in the literature to exemplify situations where the expectations from using a *blame shifting* debugging approach do not hold [21, Fig. 12]. The identified component is not the cause of the failure, and refining the type constraints of asks_for_1_string does not change the program location identified in the error. This breaks the *blame trail* property of blame shifting [21, p. 6]. We include this program to evaluate whether program slices are helpful in a situation where existing debugging approaches are known to be insufficient.

**Listing 5** The TypeScript source code for Program 6.

```
1  var functions:any = {}
2
3  function setup() {
4     functions.x = (x: number) => Math.pow(x,2)
5  }
6
7  function thrice(callback: (x: string) => string) : (s: string) => string {
8    return function(s: string): string {
9      return callback(s).repeat(3)
10   }
11 }
12
13 setup()
14
15 function action(y: string) : string {
16     return thrice(functions.x)(y)
17 }
18
19 console.assert(
20    action("hello") === "25 25 25 ",
21   "Thrice the length squared"
22 )
```

### 5.4 Program 6: Thrice

Listing 5 presents Program 6, which exhibits a more complex inconsistency where the failure location misleads from the original incorrect location. Program 6 uses first-class functions with higher-order types repeatedly as callbacks. The function thrice() produces a function that applies a stored callback to produce a string (Line 8), which is then repeated 3 times (Line 9). The function thrice() itself is correct, but its use in the body of action (Line 16) passes a function of the wrong type (functions.x has type number to number). Unfortunately, because functions has been declared in the first line as having type any, the type checker can only see functions.x as having type any and accepts the program.

## 6 Results

In this section, we describe the data obtained from the study and propose some hypotheses and future research directions.

### 6.1 Participants

The pre-screening survey filtered candidate participants for the study. Out of Prolific's pool of 2,215 participants with software industry experience, 259 participants filled our survey. Within this group, 60 answered that they had written at least 1,000 lines of code in TypeScript, but 8 of them could not clearly identify the only valid TypeScript program among the 4 options provided. From the other 52 candidates, only 11 provided the completely correct set of answers for the type constraints in our example program.

We collected 180 individual tasks from the 30 participants that finished the study. Only in 11 of these tasks (6 %) did participants use the whole assigned time and were forced to move on to the next task after 15 minutes. We discuss our process to reach 30 participants from the 52 candidates that passed the first portion of the pre-screening in the next section. From Prolific, candidate participants accessed an online prototype of our interface, hosted on a University server, which combined in the same page access to the interface and instructions for the experiment. Whenever presented, the interface for TypeSlicer was instrumented to collect information and submit task information once finished. This set up allowed us to only require candidates to have access to a web browser.

### 6.2 Pre-Screening Woes

The pre-screening program included subtleties with TypeScript types that went undetected by many participants when presented with a plain text program, but would have been easily visible in any development environment. Most subtleties arose around the treatment of `any` as a type. First, when local variables in a TypeScript program are not annotated with a type, the type checker infers their type. This is not the case for unannotated function parameters, which default to type `any`. Developers may forget this distinction and expect the type of function parameters to also be inferred to a more precise type. Second, any expression may be assigned a type `any`, even when the type system can infer a more precise type. Developers may then expect statements of the form "`X has type any`" to always hold, even when the compiler can infer a more precise type for `X`.

In our user study, another subtlety arose when one of the type annotations gave the identifier `person` the type `{x:string}`, representing an object with a single field `x` of type string. Amongst all participants, 17 developers marked the statement "`person has type string`" as correct, which is false. Participants might have not paid close attention to the subtle difference between `{x:string}` and `string`. However, any attempt to use the object as a string would have produced a compiler type error requiring only a minor correction.

We obtained 10 complete submissions from participants who answered all types correctly, 15 submissions from participants whose responses had issues with subtleties of `any`, and 5 more submissions from participants whose responses had the incorrect type for `person`. Although we opened the experiment to all participants that fulfilled these criteria, not every participant that completed the pre-screening responded to the invitation to complete the study.

### 6.3 Effect of Program Slices on Developer Focus

The main portion of the study consists of participants highlighting every part of the code that needs to be changed and providing a change recommendation. We have

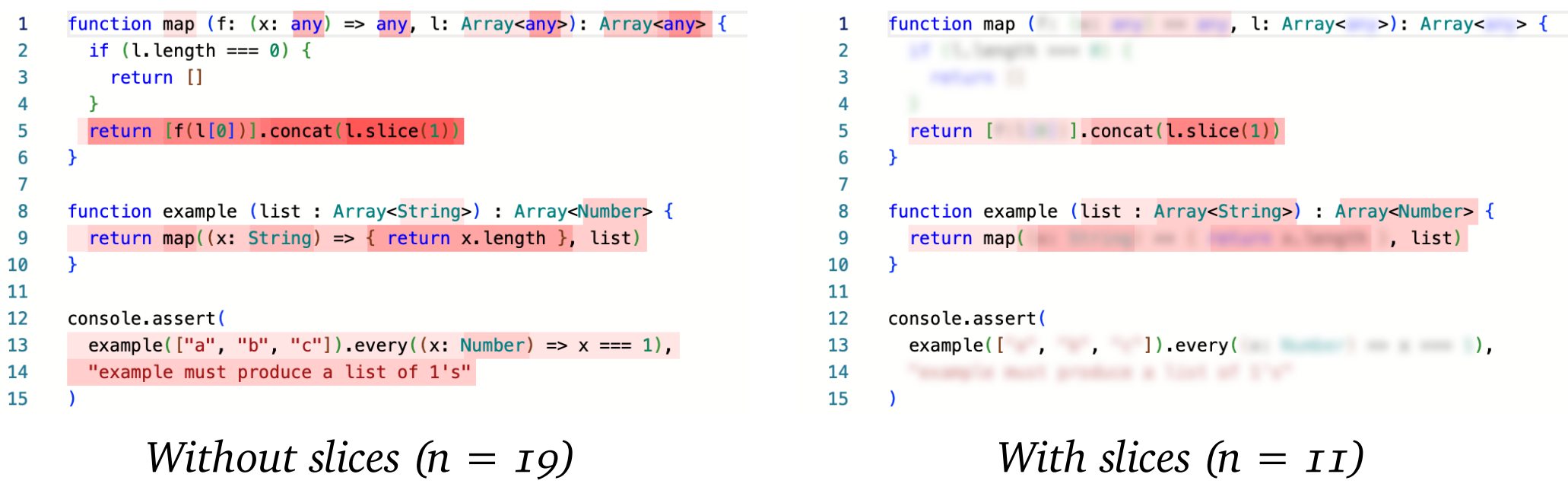

■ **Figure 2** Heat maps of participant highlights while debugging Program 1 in the absence of program slices (left), and when available from TypeSlicer (right).

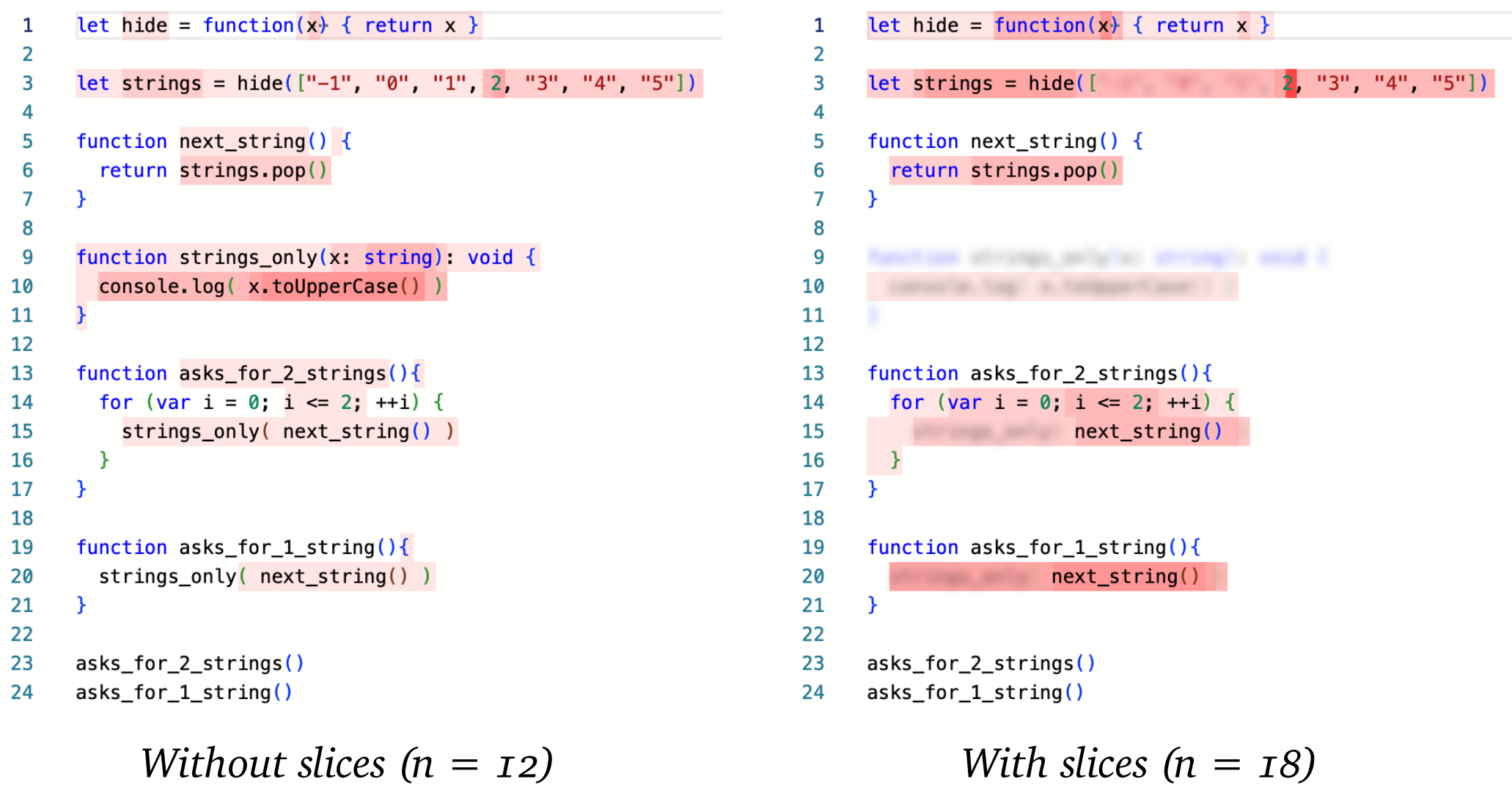

■ **Figure 3** Heat maps of participant highlights while debugging Program 3 in the absence of program slices (left), and when available from TypeSlicer (right).

aggregated all the highlights for each program as a heat map in Figures 2 to 5, where the highlighting intensity correlates with the number of participants that highlighted that section of the program. Each figure shows the highlights made by participants that had access to TypeSlicer (right panel) separately from those who did not (left panel).

For Program 1 in Figure 2 —and Program 2 and Program 4 in the appendix in Figures 8 and 9) —, highlights tended to concentrate in fewer parts of the program when TypeSlicer was available, suggesting that slices can help during debugging by narrowing down the places where developers focus. However, this concentration did not arise consistently across programs. Sometimes highlights were more spread out instead of more concentrated, as in Figure 3. In other cases, the program parts most highlighted by participants differed, as in Figure 5, suggesting a shift in focus dependent on the presence of the tool.

For Program 3 in Figure 3, 36 % of the participants picked Line 10 as an issue when TypeSlicer was not available. When TypeSlicer was available, only 6 % picked Line 10,

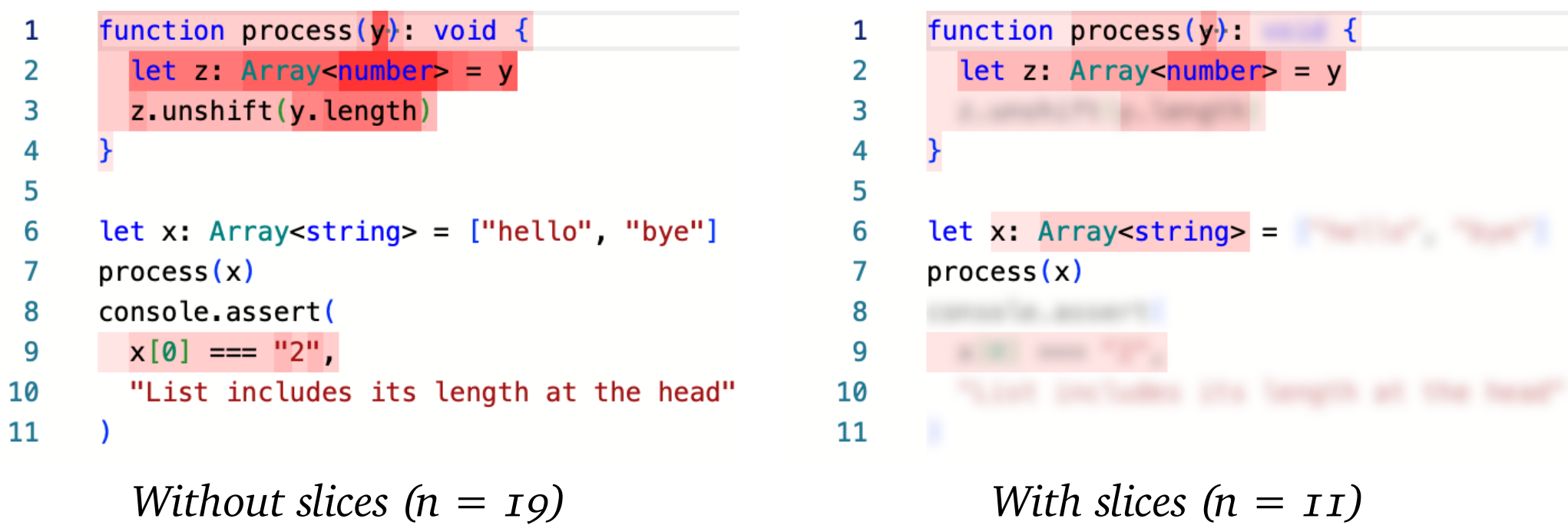

■ **Figure 4** Heat maps of participant highlights while debugging Program 5 in the absence of program slices (left), and when available from TypeSlicer (right).

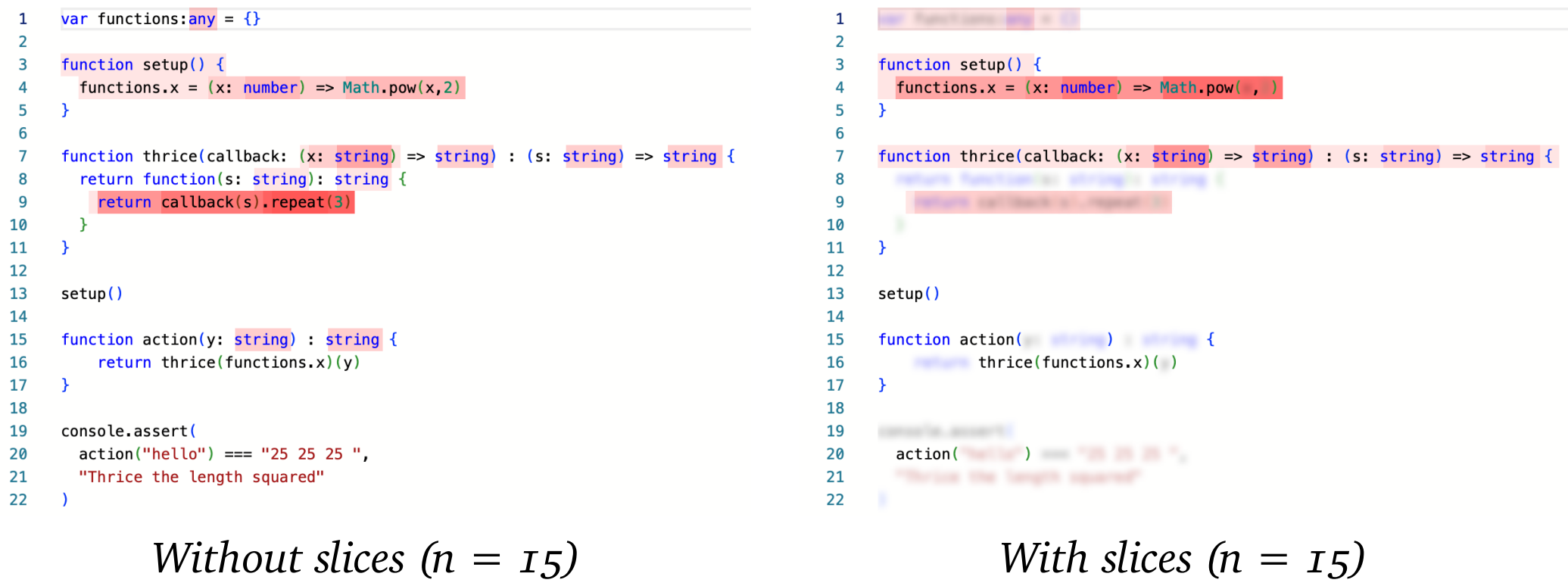

■ **Figure 5** Heat maps of participant highlights while debugging Program 6 in the absence of program slices (left), and when available from TypeSlicer (right).

while 78 % of the participants were inclined to change the contents of the strings array (Line 3). This change of focus suggests that the availability of program slices affects how participants understood the given program. Consequently, it also affected what they considered correct and incorrect: the changes on Line 9 and 10 are consistent with assuming that the value strings is the correct parameter to use inside the function, while the changes on Line 3 are consistent with assuming that the type annotation for strings_only is correct. Some of the highlights in Line 10 correspond to participants that repeatedly toggled the tool on or off (11 times), a situation that also arises for some of the highlights under the blurred portion for Program 1. But some other blurred-portion highlights are done by participants that rapidly turned the tool off, so it is hard to make any conclusions about those highlights. Making highlights on the blurred section might be more a signal of confusion about the programs themselves rather than insights about the tool, but any analysis would require further investigation.

For Program 6, Figure 5 shows similar results. Although the implementation of thrice is consistent with its type annotation, most participants without access to TypeSlicer proposed changing the implementation of thrice, which is where the bug immediately manifests. On the other hand, most participants with access to TypeSlicer proposed changing the implementation of functions.x, the first argument passed to thrice, which does not have the type signature expected for the argument of thrice.

The changes that participants recommended support this distinction. Without TypeSlicer, many participants assumed that functions.x was correct. Therefore, they suggested changes to the implementation (and signature) of thrice (80 %) to deal with numbers. For example, some participants suggested changes such as "* 3" instead of ".repeat(3)", or provided comments such as "repeat doesn't work with numbers, only strings". Participants acted in this way despite the lack of type constraints on the object functions (i.e., it has type any). Moreover, both the assertion and most other type declarations mention the type string. These responses show that, in some situations, developers side with implementation details in spite of type annotations, and may assume the annotations in the program to be incorrect. On the other hand, when slices were available, 57 % of participants suggested instead changing functions.x to produce a string.

From the highlighting data, we make the following qualitative observations:

> RQ1: When using TypeSlicer to debug a run-time type error, developers focus on different parts of the underlying program.
> RQ3: The program slices presented by TypeSlicer alter the perceptions of developers as to what changes they need to make to a program.

### 6.4 Participant Feedback

At the end of the study, participants graded TypeSlicer with an average of 75.7 (±14.4) points on the System Usability Scale. For comparison, Bangor et al. [4] report an average score of 70.14 over 2,324 surveys collected in many contexts in the span of a decade. According to Bangor et al., our average score is considered "Good", with the caveat that "scores in the 70s and 80s, although promising, do not guarantee high acceptability in the field". While the average score of TypeSlicer is not perfect, it is sufficient for the purpose of our study: TypeSlicer itself is not unacceptably bad as to derail the other reported results.

This overall positive sentiment is also supported by participant comments. One participant commented that "Typeslicer helped see where the problem is. I liked how it stopped half way through the array to see where the problem happened at the int". Another participant commented that "it was quite helpful as in task number 6 [(problem 2)] it showed me only the lines which should be corrected and after checking the code, I saw it made a good decision". A third participant also noted that "I had really big issues with finding the error without the Slicer. But, with the slicer it was 100x faster, the way it works, blurring out what it's not important really helps to focus on the problem and try to find the solution".

Participants explicitly mentioned that some of the programs given in the study are hard to understand or debug. Since we selected more complex programs for our study than other related studies [37], these answers were not surprising. However, the presence of TypeSlicer has made a difference in the perceived task complexity. Some participants with access to TypeSlicer for Program 6 mentioned that the program "seems unnecessarily complex" or "Kind of verbose". But participants who did not have access to TypeSlicer for that program gave more negative comments, e.g., "Too convoluted", or "Honestly I found this very challenging".

Participants expressed a desire to use TypeSlicer when it was missing from a particular task. One participant explicitly mentioned twice that the task was harder and that TypeSlicer would have helped them. Another participant commented while debugging Program 1 that "it took slightly longer without the TypeSlicer tool". This feedback suggests that a slicing tool, such as TypeSlicer, is helpful in debugging more complex type inconsistencies.

Although we invited participants to provide comments as part of each task, most participants provided very few comments on individual tasks aside from the required change recommendation. In general, no participant mentioned anything surprising about the output of TypeSlicer for any task, only commenting about the program behaviour and the user interface. Nevertheless, by asking for feedback per individual task first, we made participants aware that the study was looking for open feedback, and we believe this helped generate the extensive positive feedback received in the exit survey and discussed in the next subsection.

### 6.5 Program Slice to Code and Back

Did study participants make use of the slicing information, or did they ignore the slice interface? Analyzing the data shown in Figure 6 and Figure 7 helps us address RQ2. The histogram in Figure 6 shows how many times participants toggled the slice information on/off per task when TypeSlicer was available. When TypeSlicer is available for a task, it provides its feedback automatically. Therefore, zero clicks on the histogram indicates participants who only saw the program slice and never looked at the complete program. In almost a third of the task runs where TypeSlicer was available (30 % out of 90 tasks), participants never toggled the slice information, indicating that the outcome of TypeSlicer was sufficient for them to finish the task.

In 20 % of all tasks, participants at some point turned the slice information off and did not reactivate it. About half of those cases (11.1 % of all tasks) turned TypeSlicer off in the first 15 seconds after running the code, which may indicate cases where participants were not interested in the results of TypeSlicer. At the same time, in the remaining 50 % of tasks, participants went back and forth between the slice and the code in varying amounts, with an average 3.5 (±4.7) toggles per task.

The data from Figure 6 suggests that participants followed varied approaches when using TypeSlicer. The lack of toggling signals either that some participants did not realize that they could toggle TypeSlicer on/off, or that TypeSlicer always provided sufficient information. Repeated toggling might be a consequence of exploring the interface as one participant indicated "It's an interesting tool to use - I kept toggling it on and off when it was available so it was a bit distracting". Since TypeSlicer blurs the code marked as irrelevant in the slice, toggling it could also be considered as a proxy for switching attention between the slice and the rest of the program, or as a way for participants to look for extra context on programs they are unfamiliar with.

To see how TypeSlicer usage evolved during tasks, Figure 7 shows, in orange, the proportion of the running tasks where the participant has the tool on as it evolves through time since the task begins. This figure shows proportion instead of totals because participants may choose to finish a task before the 15 minute timeout, thus the total will always trend towards zero as time passes. We have additionally plotted in a blue dashed line the total of tasks where the tool was available that have been already finished after x minutes from the start of the task.

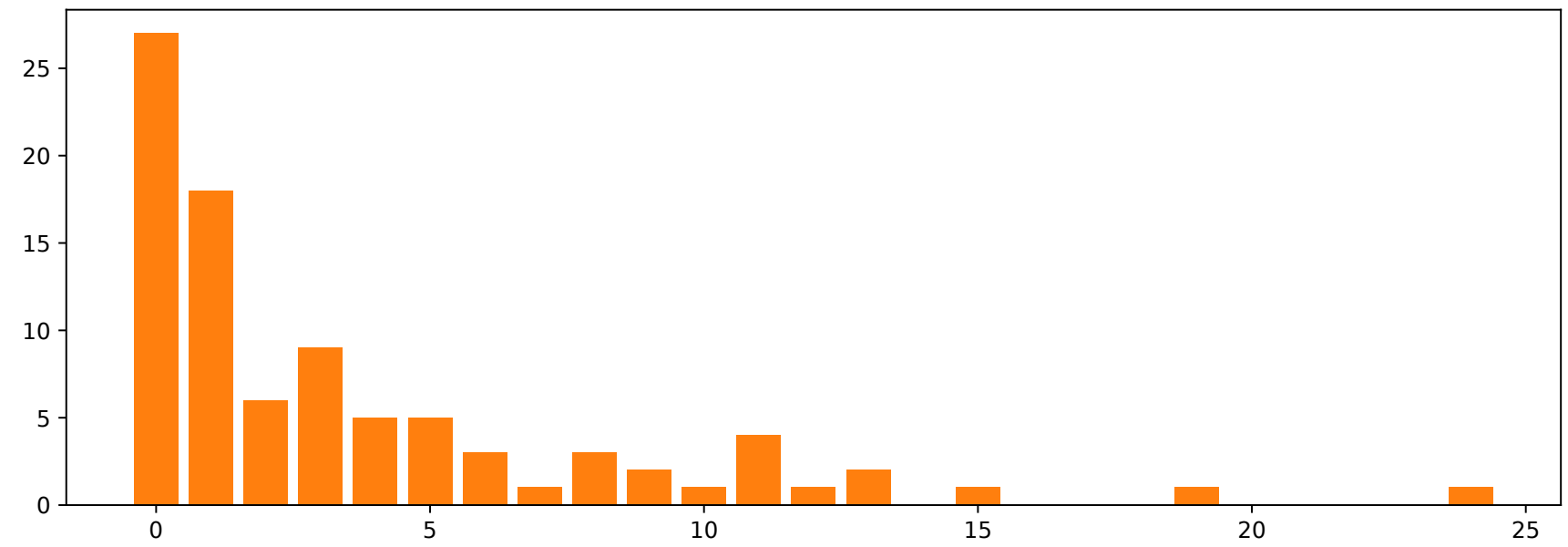

**Figure 6** Histogram of total toggles of the TypeSlicer tool button per task instance. The x-axis represents the toggle total in a particular task, and the y-axis counts the amount of task instances where that toggle total happened.

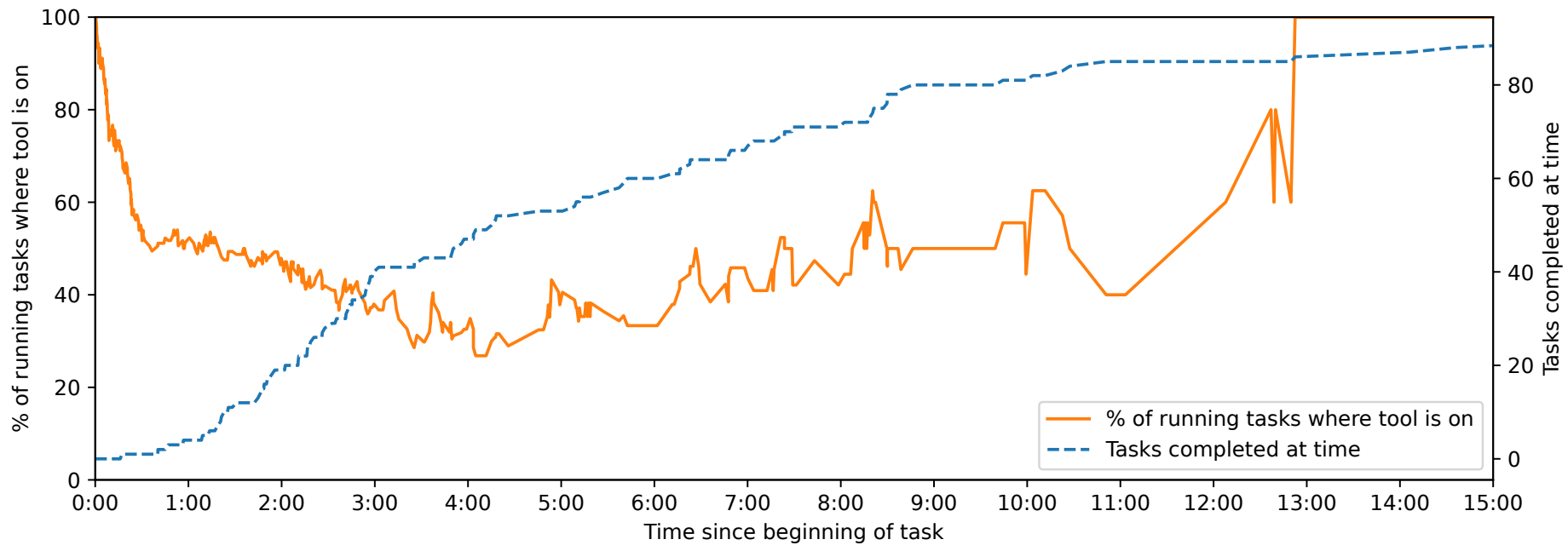

**Figure 7** Evolution of tool activation through a task: Percentage of running tasks where tool is on, and evolution of task completion.

Although the tool usage seems to decrease sharply at first, it does not do so monotonically: Our cohort of participants goes back and forth between focusing on the slice and looking at the rest of the program for context. As time passes and the number of finished tasks increases, the proportion of tasks where participants are looking at the slice information increases.

RQ2: Even in the limited context of a wizard-of-oz study, participants found slices useful and showed interest in incorporating slicing to their programming capabilities.

Overall feedback about TypeSlicer is positive and indicates that participants found slices useful. One participant said, "TypeSlicer has definitely helped me pay attention to where and why errors occur in code.". Participants recurrently showed interest in using the tool beyond the experiment, with comments like, "The TypeSlicer helped me in getting clues about where the bugs where [sic] in the programs I was debugging. It is a very nifty tool that I would like to use", "I think it's a very useful tool and can improve the productivity when working with TypeScript projects", and "I found it to be helpful to some degree and the sort of thing I would use in real life".

Not every participant comment was positive. In some situations, participants prefer accessing the slice and the code together, which TypeSlicer did not directly accommodate. Negative comments include that the tool would be improved by highlighting or underlining instead of blurring the code. Some of the negative feedback include comments such as, "It was decent, but sometimes blurring the code just made it harder to debug" and "I don't think I would use it. I mean, it could be pretty useful if it would markup code instead of blurring the rest". On the other hand, another participant stated that "By blurring individual pieces of code, TypeSlicer definitely made debugging easier". Future work should explore alternative ways to present slicing information to developers as they work.

## 7 Threats to Validity

We have identified the following threats to the validity of our results.

### 7.1 Evaluating TypeSlicer instead of Program Slices

Our study can only evaluate whether developers use program slices through a particular implementation and tool design. We limited the impact of this threat by using the System Usability Scale evaluation [5]. In (the unrealized) case of negative results, where participants could not make use of the slicing facilities at all, the SUS score could clarify whether the *user interface* was the problem instead of *slices* themselves. We applied a standardized usability tool as a preemptive check for the possibility of generalized negative feedback. However, our study found positive feedback for TypeSlicer, so this distinction was not required. Since TypeSlicer depends on slices to work and participants were able to use TypeSlicer to provide change recommendations, the study provides anecdotal support for the use of slices in this debugging context. The SUS results for the interface also support that, at the least, our interface was *good enough*.

### 7.2 Selecting Subject Programs

**Generalization to real world errors** The programs we selected are simple and somewhat unrealistic. We focused on providing participants with several programs that they could debug in short sessions with little extra context and little familiarity, attempting to include as many interesting debugging situations within the constraints of a single time-limited session. We attempted to make the programming environment close to a real-world system by reusing off-the-shelf tools (i.e. the Monaco text editor and its IDE features) and by choosing a popular programming language (i.e., TypeScript) while avoiding dependencies on language quirks and advanced typing features in the choice of our subject programs. It would be interesting to further study the interactions between program slices, language quirks, and advanced features. Nevertheless, the Wizard-of-Oz approach that we followed provides anecdotal evidence of the initial perceptions and reactions to TypeSlicer by our study participants, motivating further development that could support a more realistic long-term evaluation of the use of program slicing by developers in the gradual context.

**Rare errors** We selected subject programs that contain bugs where program slicing is potentially useful, which may represent only a fraction of all errors and bugs in gradually typed programs. Although we have shown that program slicing appears to be helpful to developers, future research should explore how often the use cases for slicing arise in real-world codebases. The positive feedback from our study participants makes us optimistic that TypeSlicer would be useful beyond the limited programs and errors situations presented in the debugging tasks of our study.

**Program size** Before checking if program slices are always useful to developers, we first wanted to check if they are useful at all. Therefore, we limited our study to a single session where we tested multiple programs. Subjecting participants to larger programs presented a time and attention trade-off against having them debug several programs in that session. Further research is required to evaluate whether program slices scale to larger programs and if there are empirical bounds to their practicality.

### 7.3 Participant Threats to Validity

Dell et al. [9] report that when participants perceive the study examiners to also be the authors of the tool under study, they tend to provide more positive feedback. To limit participant bias, we have designed the study to be performed online, without direct interaction with the tool designers. We have also carefully worded the study instructions to limit the chances of assuming a correlation between study designers and tool designers. Another threat is our sample size of 30 participants. Given our participant pool size, we do not make statistical inferences from our data, but report instead on the feasibility of slicing as a debugging tool in this particular context. In particular, this study performs no Type II error control: statistics provided are purely descriptive of the participant pool. Our sample size is comparable to user studies in HCI, as Caine [7] notes that 70 % of the user studies published at CHI 2014 reported sample sizes smaller than 30.

### 7.4 Lack of a Direct Comparison to Blame

We have not compared our approach to other state-of-the-art error approaches (e.g., blame), which may be a considerable limitation of our study. However, the main goal of our study is validating the usefulness of program slices on their own, which is a first step towards evaluation and comparison with other error reporting approaches in the presence of run-time type errors. Our methodology could also be used to evaluate blame reports in the future. With a full implementation of our approach in TypeScript, one could also compare between TypeSlicer and blame reports.

## 8 Related Work

Much work explores the formal design of gradually typed languages [31, 35] and their error semantics, but we focus on the related literature for the empirical evaluation of gradual typing and type errors.

Hanenberg et al. [17] compared users programming in statically typed and dynamically typed languages. Tunnell Wilson et al. [37] compared evaluation semantics appearing in gradual typing with a survey, and showed that developers expect the type annotations to impact the semantic behaviour of their programs. Our slicing approach would rely on a language semantics that respects declared types as supported by this research.

*Blame* [13, 35, 38] is the standard technique to report run-time type errors in gradual typing. Sometimes these blame error messages can point to places in a program that are not incorrect. Program slicing takes a different approach. Instead of pointing a single location as a cause of a failure, slices mark parts of the program as guaranteed to not be involved in the failure, an approach more resilient to situations where a single location is impossible to pinpoint or can be misleading. According to Lazarek et al. [20, 21], developers in a gradually typed language using blame should recognize when the code targeted by an error message is correct in spite of the error and impose more precise types in the program, which forces the system to assign blame somewhere else. Fixing errors thus becomes a process of increasing specification to avoid being blamed, and considerable changes to the types might be required to identify the actual cause of a blame failure. We believe slicing can be a complementary debugging approach to blame, especially in situations where blame produces misleading error messages.

Empirical evaluation of blame is limited. Lazarek et al. [20, 21] use mutation testing to evaluate whether blame errors and their debugging process leads to the actual place where the bug occurs in the program, but their approach relies on a model of a "rational programmer" which needs to be validated in practice. The main goal of their approach is to find counterexamples where the expected debugging process does not lead to the problem in a program. The existence of these situations suggests a space for alternative error debugging approaches, and highlights the need for further empirical evaluation. Future work should compare how developers approach debugging gradual programs using blame and using program slicing.

Seidel et al. [29] use a different notion of blame to justify static type error messages for type inference, proposing a strategy to synthesize a blame semantics from previous programmer errors, as "programmers eventually fix their own ill-typed programs". They very successfully predict error locations for students, but their predictions are not always correct and might not be sufficient to identify the causes of errors unseen in the training set. Program slicing takes an opposite approach, avoiding false positives for errors.

According to Weiser [40], expert programmers mentally apply program slices to reason backwards about the causes of failures, constructing in their minds an abstract representation of the programs they debug. Program slicing has been previously applied to type checking [16, 34], but not to the runtime type-errors that arise in gradual typing.

## 9 Conclusion

To reap the benefits of gradual typing in a programming language, we must pay the cost of extra run-time type errors. Debugging these errors is a challenging process due to the cost associated with identifying the root cause of the type error. Slicing provides an alternative approach for debugging run-time type errors, in a way that participants in our study found useful.

We asked participants to highlight code that should be changed, and we recorded when they toggled between seeing the whole program and only the program slice. We summarize the change suggestion highlights as heat maps (Figures 2 to 5, 8 and 9). Participant highlights tend to differ depending on availability of the tool. The highlights are sometimes concentrated in a smaller section of the program, which may indicate more focused debugging. However, the effect is not consistently observed across all programs in the study.

We summarize tool on/off toggles in a histogram (Figure 6) and a visualization of the evolution of tool activation status through tasks (Figure 7). In almost a third of the tasks, participants relied solely on the information provided in the slice, even though the interface allowed participants to choose to see the rest of the program. Still, other participants went back and forth up to 24 times to compare the full program and what was marked as irrelevant by the slice.

Although our results are limited by the very nature of a Wizard-of-Oz study, they support the development of slicing tools for gradually typed languages. Having a prototype tool that could be deployed and used with arbitrary programs would provide users with the opportunity to apply slicing in their day-to-day work and in codebases that they are already familiar with. This prototype implementation would allow the evaluation of TypeSlicer among professional developers in a real-world context.

Our user study shows that TypeSlicer was sufficiently usable for our participants who gave the tool an above-average System Usability Scale score of 75.7. Participant have also provided many positive feedback comments.

To generalize our results, more research is required to improve tool usability, and explore hypotheses from our observations. For example, alternatives to blurring the sliced-out portions of a program would alleviate most of the negative reactions we observed. Nevertheless, our study provides initial empirical evidence that slices are useful, justifying future evaluation and design of slicing systems and their semantics for gradually typed languages.

**Listing 6** The TypeScript source code for Program 2.

```
1  let container = wrap_again( () => "2OOO")
2
3  function wrap (x: () => number): () => number {
4    return x
5  }
6
7  function wrap_again (x) : () => number {
8    return wrap(x)
9  }
10
11 function main (f: () => number): bigint {
12   return BigInt( f()*10000 )
13 }
14
15 console.assert(
16   19999999<= main(container),
17   "result is big enough"
18 )
```

# A Discussion of Programs 2 and 4

For space reasons, we discuss programs 2 and 4 in the appendix, following the same structure of the rest of the paper.

## A.1 Program 2: Wrap

Listing 6 presents Program 2, where the wrap_again function does not statically check the type of its argument. This happens because any function parameter without a type declaration defaults to any in TypeScript. The call to wrap_again on line 1 manages to hide the type of a function that returns a string and present it to the rest of the program as a function returning a number that is stored in the container variable.[1] The problem arises when multiplying the result of f() on line 12, but the problem is far removed in a set of stacked type wrappers.

[1] In some programming languages, this type difference would be sufficient to make the program fail. But TypeScript, as an extension of JavaScript, relies on implicit conversions and could convert the string "2000" to the number 2000 whenever needed and continue evaluation. This behaviour can lead to unexpected results beyond the scope of this study. To ensure this program always fails in TypeScript, we go a step further and replace the zeros in the string with the capital O letter. The string "2OOO" in the first line cannot then be implicitly converted by TypeScript into a number.

**Listing 7** The TypeScript source code for Program 4.

```
1 let t: any = ["A", 3]
2 let strs: Array<string> = t
3 let fst: string = strs[1]
4 fst.toUpperCase()
```

This program is a simplification of an example program identified by Lazarek, King, Sundar, Findler, and Dimoulas [21] (Fig. 13, a mutation of KCFA), which they find interesting for the evaluation of blame tracking with contracts.[2] The program represents a set of nested type constraints where the path of inconsistency is somewhat complex, and we expected slices to draw participants' attention away from the implementation of `main` and towards the argument to `container`.

### A.2 Program 4: Non-Uniform List

Listing 7 presents Program 4, a simple program that is a variation on the "Program 2" proposed by Tunnell Wilson, Greenman, Pombrio, and Krishnamurthi [37] (not our program 2). The list `t` is not uniform, and the element selected of the list, `strs[1]`, is a number and not a string. Although the first three lines should suffice to make an error in a type safe language, they do not suffice in TypeScript. The fourth line was added to guarantee that the program produces an error when the type safety checking tool is not in use.

This is a simple example without any branching and where the type annotations do not correspond to the values hiding behind a list of type `any`. Tunnell Wilson, Greenman, Pombrio, and Krishnamurthi [37] identified that some developers expected the program to fail on line 2, where the non-uniform array was cast to an array of strings instead of when the array is accessed. This suggests *"that developers prefer a strategy that type-checks the elements of an untyped array immediately when the array flows into a typed context instead of waiting until the array is accessed."* [37]. We expect this program to provide an easier and short example through the experiment and that the highlights will vary among participants on where to change the program.

[2] The failure of interest in the original publication relies on lazy contract enforcement for mutable references in Racket (`box/c`). The lazy contract enforcement of `box/c` impacts the contract checking order, and that breaks one of the assumptions for the "blame shifting" debugging technique. We replace the `box/c` contracts with higher order types, as the original problem does not arise in the formalism that we use to generate the slices, which is discussed in detail in the first author's dissertation [1].

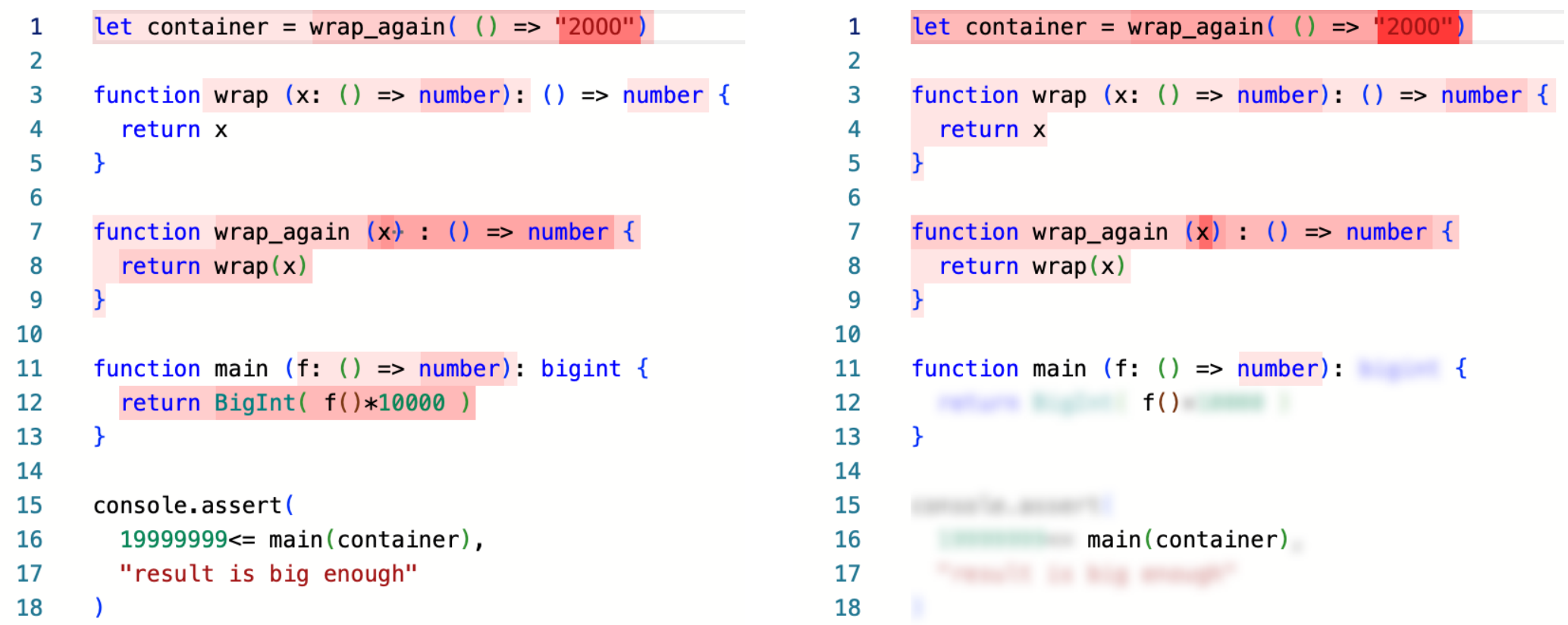

*Without slices (n = 12)* *With slices (n = 18)*

■ **Figure 8** Heat maps of participant highlights while debugging Program 2 in the absence of program slices (left), and when available from TypeSlicer (right).

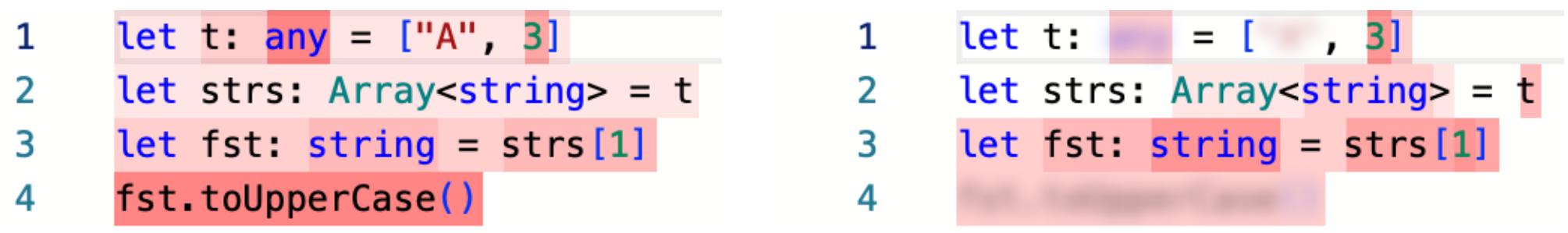

*Without slices (n = 13)* *With slices (n = 17)*

■ **Figure 9** Heat maps of participant highlights while debugging Program 4 in the absence of program slices (left), and when available from TypeSlicer (right).

## About the authors

**Felipe Bañados Schwerter** banadoss@ualberta.ca
https://orcid.org/0009-0001-1879-8482

**Ronald Garcia** rxg@cs.ubc.ca
https://orcid.org/0000-0002-0982-1118

**Reid Holmes** rtholmes@cs.ubc.ca
https://orcid.org/0000-0003-4213-494X

**Karim Ali** karim.ali@nyu.edu
https://orcid.org/0000-0002-5516-1376